\documentclass{kluwer}    
\usepackage{psfig}

\def\lhls{${\cal L}_h/{\cal L}_s$}

\newdisplay{guess}{Conjecture}

\begin{document}
\begin{article}
\begin{opening}
\title{Accretion in Strong Gravity: From Galactic to Supermassive Black Holes}
\author{Chris \surname{Done} and Marek \surname{Gierli\'nski}}
\runningauthor{Chris Done and Marek Gierli\'nski}
\runningtitle{Black Holes and Strong Gravity}
\institute{Department of Physics, University of Durham, UK}
\date{}

\begin{abstract}

The galactic black hole binary systems give an observational template
showing how the accretion flow changes as a function of increasing mass
accretion rate, or $L/L_{Edd}$. These data can be synthetised with
theoretical models of the accretion flow to give a coherent picture of
accretion in strong gravity, in which the major hard-soft spectral
transition is triggered by a change in the nature and geometry of the
inner accretion flow from a hot, optically thin plasma to a cool,
optically thick accretion disc.  However, a straightforward application
of these models to AGN gives clear discrepancies in overall spectral
shape. Either the underlying accretion model is wrong, despite its
success in describing the Galactic systems and/or there is additional
physics which breaks the simple scaling from stellar to supermassive
black holes.

\end{abstract}
\keywords{black holes, accretion flows}

\end{opening}

\section{Introduction}

The famous quote by John Wheeler that ''Black holes have no hair''
refers to their amazing simplicity. Theoretically they can be {\em
completely} described by mass, spin and charge, while in any realistic
astrophysical situation this reduces to simply mass and spin. However,
black holes are most easily studied if they accrete, where the
infalling material converts some of its immense gravitational potential
energy to high energy radiation before disappearing forever below the
event horizon. Thus there is another parameter which describes the
appearance of the most easily observed black holes, namely their mass
accretion rate.

This theoretical simplicity is at first glance wildly at odds with the
observed complexity of emission from accreting black holes. This is
especially evident in the stellar mass black holes in our Galaxy
(GBHC), where there is now a huge amount of high signal-to-noise data
covering a large range of different mass accretion rates. However,
recent progress has shown that these data can all be fit together into
a coherent phenomenological framework, and that this can plausibly
relate to physically based models of the accretion flow (Done \&
Gierli{\'n}ski 2003, hereafter DG03). The general picture emerging from
the data is that the major hard-soft transition seen in the GBHC is
consistent with being triggered by a change in the nature and geometry
of the inner accretion flow from a hot, optically thin, geometrically
thick plasma to a cool, optically thick, geometrically thin disc
(Poutanen et al 1997; Esin et al 1997).

Here we try to scale up the physical models of accretion which are so
successful in describing the data from the Galactic black holes to the
accreting supermassive black holes which power Active Galactic Nuclei
(AGN) and Quasars.  The goal is synthetise both theory and
observations, to build a physically based model which can explain the
data from accretion flows onto all masses of black hole.

\section{Galactic Black Hole Binary Systems}

The GBHC all have fairly similar mass, but show a wide variety of mass
accretion rates due to the disc instability (King \& Ritter
1998). These data give a observational template showing how the
accretion flow varies as a function of (predominantly) mass accretion
rate i.e. $L/L_{Edd}$. The standard disc models predict a very robust
quasi-blackbody spectrum, with temperature $kT_{disc} \sim 1
(M/M_\odot)^{-1/4} (L/L_{Edd})^{1/4}$~keV i.e. $\sim 1$~keV for a
$10M_\odot$ GBHC accreting at the Eddington limit.  Such spectra are
seen, but are generally accompanied by a weak (ultrasoft state: US),
moderate (high state: HS) or strong (very high state: VHS) X-ray tail
to higher energies. Together these form the soft states, which are
seen at high $L/L_{Edd}$. However, at low $L/L_{Edd}$ these objects
can also show spectra which look entirely unlike a disk, peaking
instead at $\sim 100$ keV (low/hard states: LS). Fig. 1a shows
representative spectra from all these GBH states (e.g. the review by
Tanaka \& Lewin 1995).

\begin{figure}
\begin{tabular}{cc}
\psfig{file=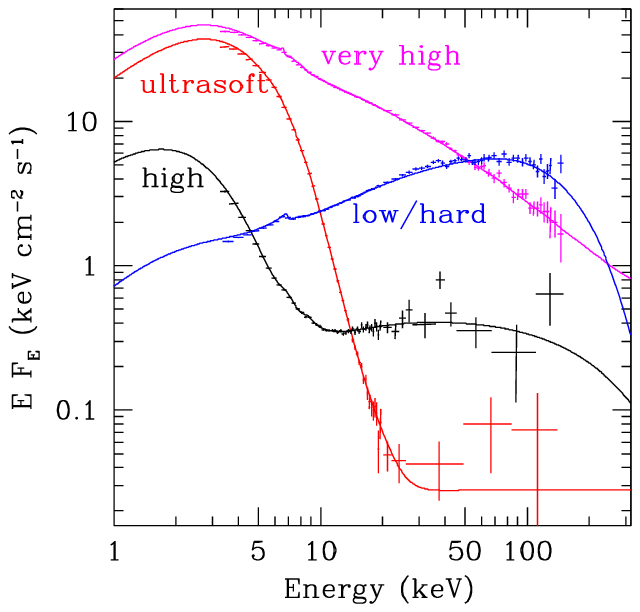,width=0.35\textwidth} &
\psfig{file=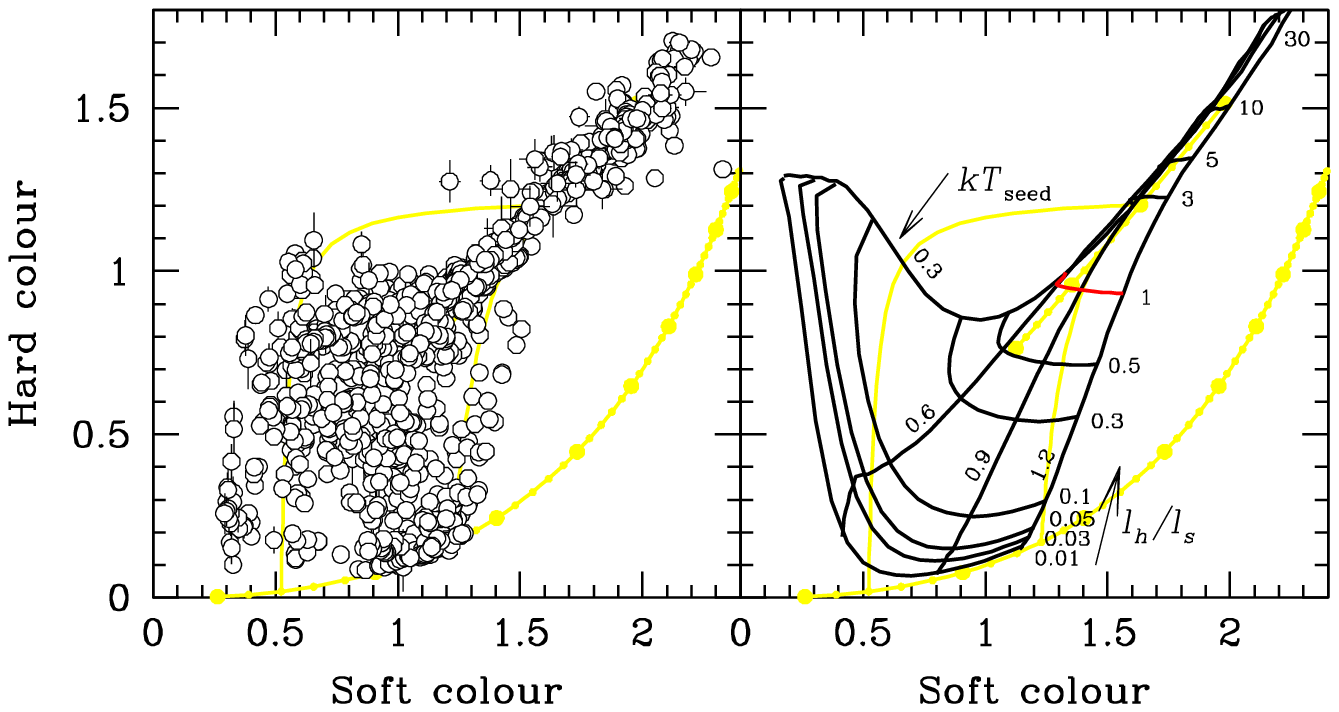,width=0.60\textwidth}\\
\end{tabular}
\caption[]{The left panel shows RXTE PCA and HEXTE data from
XTE~J1550-564 to illustrate the range of spectral shapes seen from the
GBH.  The middle panel compresses all the spectral information into
colours, with soft and hard colours roughly related to the inverse of
the mean spectral slope between 3--6~keV and 6--16~keV, respectively.
All the black holes are consistent with the same behaviour. The right
panel shows that this behaviour can be reproduced using Comptonisation
models constrained by the sketched geometries shown in Fig. 2.}
\label{geom}
\end{figure}

To produce any emission at energies substantially higher than that of
the disk {\em requires} that some fraction of the gravitational energy
is dissipated in regions which are optically thin, so that a few
electrons gain a large fraction of the energy. These energetic
electrons can produce hard X-rays by Compton upscattering lower energy
photons, and the shape of this spectrum is determined by the ratio of
power in the hot electrons to that in the seed photons illuminating
them, \lhls.

While such Comptonization models can explain the broad
band spectral shapes, they do not address the underlying problem of
the {\em physical origin} of the hot electrons, or indeed the range
\lhls\ required to produce the very different spectra shown in
Fig. 1a.  We can get some insight into these more fundamental issues
from recent advances in understanding the physical nature of the
accretion disc viscosity as a magnetic dynamo (Balbus \& Hawley 1991).
Numerical simulations show that any seed magnetic field can be
continuously amplified by the differential rotation of the disc
material, and dissipated through reconnection events. Including
radiative cooling gives an accretion disc structure which bears some
resemblance to the standard accretion disc models, but with some of
the magnetic reconnection occurring above the disc as magnetic field
loops buoyantly rise to the surface, reconnecting above the bulk of
the material in an optically thin environment (e.g. Turner 2004).

However, these physical viscosity simulations also show that an
alternative, {\em non-disc} solution can exist, where the whole
accretion flow is optically thin, so cannot efficiently cool. The
accretion flow forms a hot, geometrically thick structure,
qualitatively similar to the Advection Dominated Accretion Flows
(Narayan \& Yi 1995), but considerably more complex in detail, with
convection (e.g. Igumenshchev et al 2003) and outflows (Blandford \&
Begelman 1999) as well as advection (Hawley \& Balbus 2002).

The existence of two very different accretion flow structures gives a
very natural explanation for the two very different types of spectra
(hard and soft) seen from the GBH. At low $L/L_{Edd}$ the inner
optically thick disk is replaced by an optically thin flow. There are
few photons from the disk which illuminate the flow, so \lhls$\gg1$
and the Comptonised spectra are hard. When the mass accretion rate
increases, the flow becomes optically thick, and collapses into an SS
disk. The dramatic increase in disk flux drives the hard-soft state
transition (Esin et al 1997). A weak tail on the dominant disk
emission can be produced by occasional magnetic field loops buoyantly
rising to the surface, reconnecting above the bulk of the material in
an optically thin environment (US).  Increasing the ratio of power
dissipated above the surface to that in the disk increases \lhls,
increasing the importance of the hard X-ray tail. However, the {\em
geometry} of the soft states sets a limit to \lhls. Flares {\em above}
a disk illuminate the disk surface, where some fraction are absorbed
and thermalised. This adds to the intrinsic disk emission, fixing
\lhls$\lsim 1$ in the limit where the flares cover most of the disk
surface (Haardt \& Maraschi 1993), which always results in a soft
Comptonised spectrum, forming a power law with energy index $\alpha
\gsim 1$ (VHS). Fig. 2 illustrates the geometries inferred for each
state.

\begin{figure}
\psfig{file=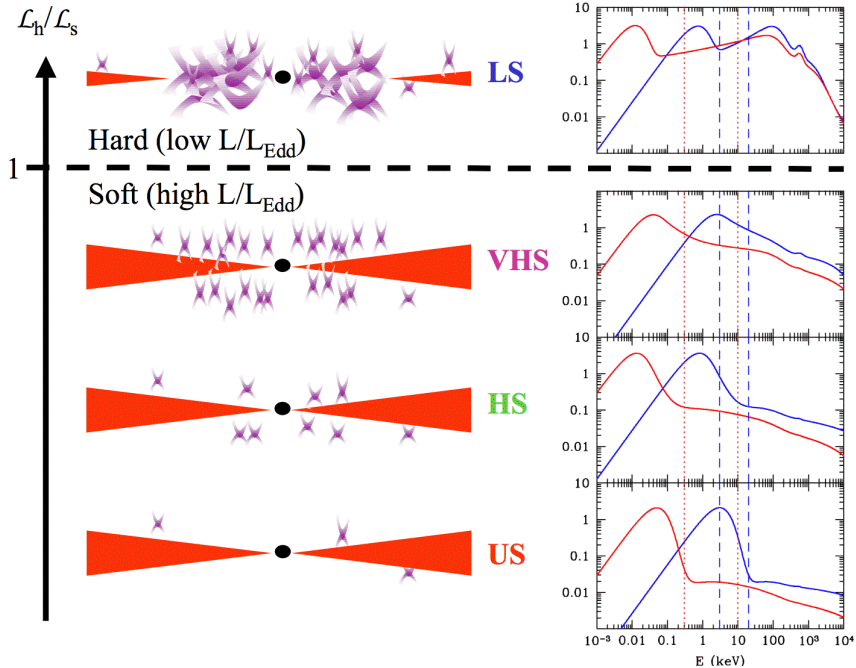,width=0.95\textwidth}
\caption[]{Sketched geometries corresponding to the spectral states,
together with the {\sc eqpair} spectra for GBH (higher disc
temperature) and AGN (lower disc temperature). The dashed and dotted
vertical lines on the spectral panels show the energy ranges for {\it
RXTE} and {\it XMM-Newton}, respectively}
\label{geom}
\end{figure}

\section{Quantative models of the X-ray spectra of GBH}

The picture developed above for the geometry of the accretion flow
puts constraints on the expected emission. The energetic electrons in
the optically thin regions can produce hard X-rays by Compton
upscattering lower energy photons, and the shape of this spectrum is
broadly determined by the ratio of power in the electrons to that in
the seed photons illuminating them, \lhls.  However, the spectral
shape of the Comptonised emission also depends to some lesser extent
on the details of the electron distribution (its optical depth and
whether it is thermal, nonthermal, or has some more complex shape) and
seed photons (temperature and spectrum).

Detailed modelling of individual spectra from GBH show that the X-ray
emission in the low/hard state it is fairly well modelled by thermal
Comptonisation of accretion disc photons by hot, thermal ($\sim
100$~keV) electrons (e.g. Gierli{\'n}ski et al. 1997), although there
may be some evidence for non-thermal electrons also being present
(McConnell et al 2002). However, in the soft states the spectral
curvature in the tail is clearly best described by a combination of low
temperature thermal ($\sim 10$ keV) {\em and} nonthermal electrons
(Gierlinski et al 1999; Zdziarski et al. 2001; Frontera et al. 2001;
McConnell et al 2002; Gierli{\'n}ski \& Done 2003; Kubota \& Done
2004). These could be two physically (and perhaps spatially) distinct
populations, or a single 'hybrid' plasma. The latter idea comes from
the fact that even a purely non-thermal acceleration process cannot
give rise to a completely power law electron distribution as
electron-electron collisions will always give rise to {\em some}
thermalisation at the lowest energies (Coppi 1999). Alternatively, even
assuming that the energy injection to the electrons is purely thermal
leads to a non-thermal tail from stochastic scattering (second order
Fermi processes) on magnetic field inhomogeneities (Dermer, Miller \&
Li 1996; Liu, Petrosian \& Melia 2004). Thus it seems very likely that
the electron distribution is indeed complex, even if we are dealing
with a single acceleration region.

Thus the simplest model for the emission is one where there is a
single acceleration process for the magnetic reconnection irrespective
of its spatial location (hot inner flow or flares above a disc). We
use the sophisticated Comptonization code, {\sc eqpair} (Coppi 1999) to
translate this schematic picture into a {\em quantitative} model. The
key advantage of this code is that it does not assume a steady state
electron distribution, rather it {\em calculates} it by balancing
heating (injection of power ${\cal L}_h$ into thermal and/or
non-thermal electrons) and cooling processes (Compton cooling, which
depends on ${\cal L}_s$, Coulomb collisions, photon-photon collisions
leading to e$^{+/-}$ pair production and annihilation). The resulting
spectrum depends primarily on \lhls, i.e. on the geometry, and on the
form of the electron injection. Guided by the results from detailed
fits to individual spectra, we choose a constant electron injection
spectrum which has optical depth of unity, with the power split
equally between non-thermal (power law of $\Gamma_{inj}=2.5$ up to
maximum Lorentz factor of $10^3$) and thermal components.

Fig. 1c shows a grid of colours resulting from the {\sc eqpair} code
for \lhls changing from 30 (top right of the diagonal branch) -- 0.01
(softest hard colours), assuming seed photons from the disc at 0.3 --
1.2 keV as expected for the observed range in $L/L_{Edd}$ for a
standard disc.  Changing only these two {\em physical} parameters can
describe {\em all} the colour evolution seen from the GBH. These model
spectra for each state are shown in the right hand panel of Fig. 2
darker line, with higher disc temperature), with the dotted lines
showing the energy range of the PCA data over which the colours are
measured (DG03).  These {\em same} models for the accretion flow (both
qualitative and quantitative) can also explain the very different
colours seen from the disc accreting neutron star systems. These have
similar gravitational fields, so should have similar accretion flows,
but with the addition of a boundary layer between the flow and the
solid surface (DG03).

\section{Application to supermassive black holes}

\begin{figure}
\begin{tabular}{cc}
\psfig{file=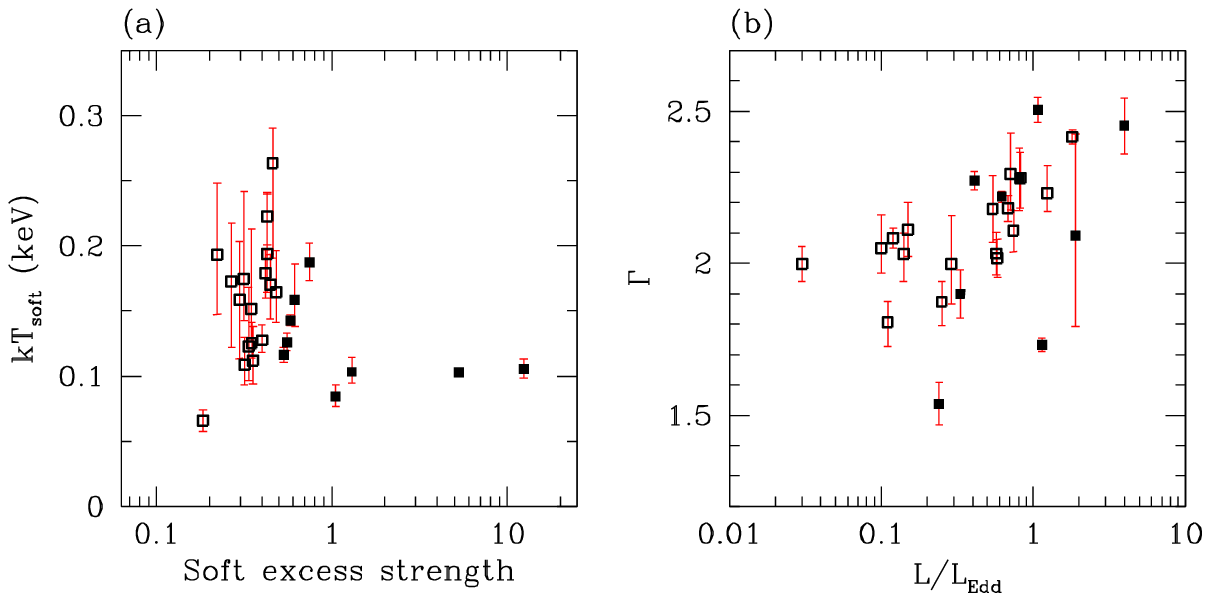,width=0.65\textwidth} &
\psfig{file=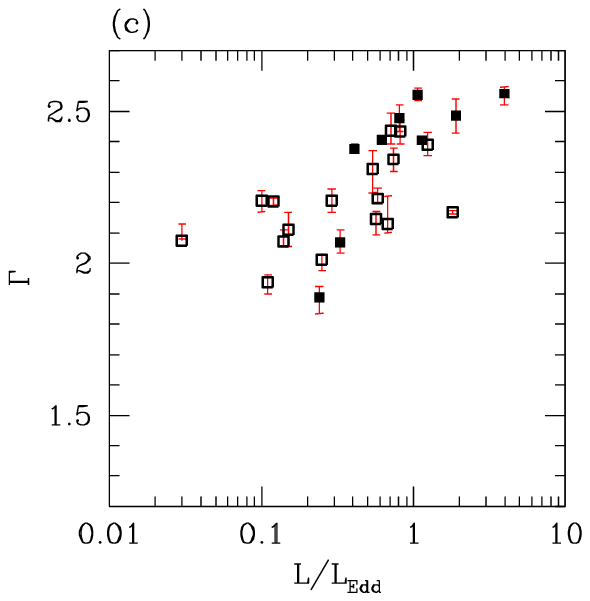,width=0.31\textwidth}\\
\end{tabular}
\caption[]{Panel (a) shows the soft excess strength versus its
temperature. All the PG quasars used here have very similar
temperatures despite large differences in mass and mass accretion rate.
Panel (b) shows the power law spectral index inferred from fits
including a soft excess. Several objects have high mass accretion rates
but rather hard spectra, in conflict with the behaviour of GBH. Panel
(c) shows this conflict is removed by using a model in which there is
complex absorption from partially ionized material with a range of
velocities along the line of sight (as expected from a wind from the
inner disc).} \label{exc}
\end{figure}

AGN accretion flows should be similar to those in GBH at the same
$L/L_{Edd}$, except that the much larger mass black hole leads to a
lower accretion disk temperature (Shakura \& Sunyaev 1973).  However,
studying the AGN accretion flows is difficult as the black hole mass
is much harder to determine, so giving large uncertainties on
$L/L_{Edd}$. This problem can now be addressed using the recently
discovered correlations of central black hole mass with the
luminosity/velocity dispersion of the bulge, or the line width of the
narrow line region, or reverberation mapping (e.g. Woo \& Urry
2002). The other problem is signal-to-noise, with only the $\sim 10$
brightest AGN having adequate spectra in RXTE. Again, this is now
changing due to the unprecedented sensitivity of the EPIC camera
(0.2--10 keV) on ESA's {\it XMM-Newton} satellite

If the models which describe the GBH spectra really do work then we
can use the same code to {\em predict} what we should see from AGN,
assuming that the {\em only} change in the accretion flow structure is
due to the mass of the black hole changing the temperature of the
disk. Since $kT_{disk}\propto M^{-1/4}$ for a given $L/L_{Edd}$
then the GBH temperature range of 0.3--1.2~keV observed in the
$10M_\odot$ stellar black holes scales to 5--20~eV for a $10^8M_\odot$
AGN.  The grey lines in the spectral panels of Fig. 2 show the effect
of this seed photon temperature change on the {\sc eqpair} Comptonised
spectra assuming the {\em same geometries} as used for the GBH.

The vertical lines on these spectra show the relevant bandpasses of
the {\it RXTE} PCA (GBHC: dashed line) and {\it XMM-Newton} EPIC
instruments (AGN: dotted), respectively. Plainly these models predict
that the analogue of the soft states in AGN will have no direct disk
emission in the XMM-Newton bandpass. The predicted spectra are
approximately power laws, and soft excesses should be weak and
rare. For all the states, the lower seed photon temperature means that
the Comptonised spectra extend to lower energies and are slightly
softer. Thus the soft state spectra always have Comptonised emission
with $\alpha>1$, and even the hard state spectra have $\alpha > 0.8$.

The bright quasar sample are objects selected by their strong blue/UV
continuum flux, i.e. have a strong accretion disc component so should
correspond to the soft state. This is confirmed by their estimated
$L/L_{\rm Edd}$, with the majority spanning the range between $0.1 <
L/L_{\rm Edd} < 1$.  We selected all the publicly available (as of
September 2003) X-ray spectra from \emph{XMM-Newton} archive.  We fit
these 26 objects with a continuum model consisting of two Comptonized
components. The hot component produces the power-law spectrum, while
the cool one gives freedom to model any soft X-ray excess.  Contrary to
expectations, {\em all} the objects {\em require} a soft excess
component. Fig 3a shows the characteristics of the soft excess for each
object, plotting temperature against strength of the soft excess,
$R_{\rm exc}$, measured by the ratio of unabsorbed 0.3--2 keV flux in
the cool and hot components.  The most striking property of the soft
excess is its constancy in temperature. It is distributed in a very
narrow range of values between 0.1 and 0.2~keV, and does not correlate
in any way with the expected disc temperature estimated from black hole
mass and $L/L_{Edd}$. Equally contrary to expectation is the spectral
index of the hot Comptonisation component. Fig 3b shows this plotted
against the estimated $L/L_{Edd}$ for each object. While there is the
same general trend as in the GBHC for high $L/L_{Edd}$ objects to be
steeper, there are several AGN which are at high Eddington fractions
which have $\alpha = \Gamma-1 < 1$, and that many of these have strong
soft X-ray excesses ($R>0.5$), denoted by filled symbols (Gierli\'nski
\& Done 2004).

PG~1211+143 is the most extreme example of this in our sample. It has
$\alpha\sim 0.8$ for the intrinsic (reflection and ionised absorption
corrected) continuum, and also has a strong soft X-ray excess.
Similar objects are also seen in the literature, with the strongest
soft excesses often seen in Narrow Line Seyfert 1's  (e.g. 1H 0419-577:
Page et al 2002; 1H 0707-495: Fabian et al 2002).  Plainly there are
problems in a simple application of the GBH spectral models to
AGN. Either the GBH models are wrong, or there are additional physical
processes which break the scaling between AGN and GBH.

\section{Additional complexity in AGN spectra ?}

One obvious candidate for additional complexity in the AGN spectra is
the generic presence of partially ionised absorption. The environment
around an AGN is often gas-rich, and X-ray illumination of distant
material such as the molecular torus can form a partially ionised wind
(Krolik \& Kriss 2001). However, changes in this absorption on fairly
short timescales suggest that at least some component of this is
directly associated with the disk (e.g.  Pounds et al 2003). The low
disk temperature in AGN means that most of the disk material has
substantial opacity from all elements except H and He. There are
multiple line transitions from these elements in the UV band, where
the disk spectra peak, so these can result in a strong line-driven
wind from the disk. By contrast, the higher disk temperature in GBH
means the disk has much lower opacity, predicting a much weaker wind
(Proga \& Kallman 2002).

The gratings on {\it XMM-Newton} and Chandra have shown the ionised
absorption in AGN in unprecedented detail.  In general, multiple
absorption components are seen, with different outflow velocities,
columns and ionisation states (e.g. Blustin et al. 2002). However,
these absorbers are included in the fits to PG~1211+143, and make no
substantial difference to the size of soft excess or hardness of the
2--10~keV spectrum. However, these absorption components are identified
by their {\em narrow} atomic features, implying that the dispersion in
velocity along the line of sight is rather small. This is {\em not}
what is expected from the disc wind described above.  Instead this
should be differentially rotating, and outflowing, so has a very
complex velocity structure which gives substantial {\em broadening}
(Murray \& Chiang 1997).

We refit the data with a model in which there is only one Comptonised
component i.e. no additional soft excess, together with a simple model
of the absorption expected from a discwind (an ionised absorber
convolved with a Gaussian velocity dispersion). Fig 3c shows the new
distribution of spectral indices with $L/L_{Edd}$. {\em All} the AGN
now have intrinsically steep spectra ($\alpha>1$) as expected for the
supermassive analogues of the soft state GBHC. The typical velocity
dispersions are $\sim 0.1-0.3c$, as expected if the wind is launched
from close to the last stable orbit of the disc, and the columns
required are $\sim 10^{21-23}$ cm$^{-2}$ (Gierli\'nski \& Done 2004).
Such absorption models models can fit the
data from individual bright objects as well as a
separate soft excess component or ionised reflection (Sobolewska \&
Done 2004).

\section{Conclusions}

{\em All} the data from the galactic black hole binaries is consistent
with showing the same spectral evolution as a function of increasing
$L/L_{Edd}$. This evolution can be qualitatively modelled by a change
in the nature and geometry of the accretion flow, from a hot,
geometrically thick plasma to a cool, geometrically thin disc.  The
implications of this on the emitted spectrum can be quantified using
sophisticated Comptonisation codes, and these can match the observed
data. These models, with the addition of a boundary layer, can also
explain the rather different spectral evolution seen from the disc
accretion neutron star binary systems.

The Comptonisation models can easily be scaled up to {\em predict} the
spectra from AGN and quasars, assuming that the physics of the
accretion flow is the same. However, these predictions conflict with
the observed spectra of the PG quasar sample. These all have high
$L/L_{Edd}$ so should be soft state analogues of the GBHC, but several
have rather hard 2--10~keV spectra, and all require an additional soft
X-ray component which has no obvious counterpart in the GBHC. Instead
we suggest that these spectra {\em are} as predicted by the models,
but that our view of them is distorted by complex, partially ionised
absorption from an accretion disc wind. The large velocity shifts in
the wind smear the intrinsically narrow absorption features so that
the material gives no clearly identifiable signal in high resolution
grating data. The difference between the GBHC and AGN is then that the
AGN have strong absorption from a discwind, while the GBHC do
not. This can easily be explained by the much lower disc temperature
expected in supermassive black holes. Firstly this means that the disc
itself retains substantial opacity, so there is much more line driving
force for launching the wind, and secondly, heavy elements in the wind
are less likely to be completely ionised, so have more effect on the
X-ray spectrum. While such models are speculative, the alternative
is that we missing some substantial piece of accretion physics in the
galactic black hole models.

\end{article}
\end{document}